\documentclass[12pt,fleqn]{article}

\usepackage[margin=1in]{geometry}
\RequirePackage[T1]{fontenc}
\usepackage{setspace}
\RequirePackage{amsthm,amsmath}
\usepackage[round]{natbib}
\usepackage[english]{babel}
\usepackage[colorlinks,citecolor=blue,urlcolor=blue]{hyperref}
\usepackage{enumerate}
\usepackage{caption}
\usepackage[english]{babel}
\usepackage{nameref}
\usepackage{prepub}
\usepackage{xcolor}
\usepackage[bbgreekl]{mathbbol}
\usepackage{comment}
\usepackage{amsmath,amsfonts,amssymb,euscript,mathrsfs,mathtools, mathdots}
\usepackage{graphicx}
\usepackage[inline]{enumitem}
\usepackage[Sonny]{fncychap}
\usepackage{bigints}
\usepackage{xcolor}
\usepackage{booktabs, multirow} 
\usepackage{soul}
\usepackage{changepage,threeparttable} 
\usepackage{fullpage}
\usepackage{algorithm}
\usepackage{algpseudocode}
\usepackage{amsmath}
\usepackage{dsfont}
\makeatletter
\def\captionof#1#2{{\def\@captype{#1}#2}}
\makeatother
\def\1{\mbox{\bf 1}}
\def\R{\mathbb{R}}

\def\N{\mathbb{N}}
\def\P{\mathbb{P}}
\def\E{\mathbb{E}}

\def\R{\mathbb{R}}

\def\Z{\mathbb{Z}}
\def\e{\mathrm{e}}

\newcommand{\dpartial}[2]{
	\frac{\partial {#1}}{\partial {#2}}
}

\def\independenT#1#2{\mathrel{\rlap{$#1#2$}\mkern2mu{#1#2}}}

\newtheorem{prop}{Proposition}

\newtheorem{Def/Prop}{Definition-Proposition}

\newcounter{ex}

\renewcommand{\theex}{\thesection.\arabic{ex}}

\newcounter{exos}
\renewcommand\theexos{\arabic{exos}}

\newcounter{prob}
\renewcommand\theprob{\arabic{prob}}

\newcounter{note}
\renewcommand{\thenote}{\thesection.\arabic{note}}

\makeatletter
\let\orgdescriptionlabel\descriptionlabel
\renewcommand*{\descriptionlabel}[1]{%
	\let\orglabel\label
	\let\label\@gobble
	\phantomsection
	\edef\@currentlabel{#1}%
	\let\label\orglabel
	\orgdescriptionlabel{#1}%
}
\makeatother

\begin{document}

\date{~} 

\title{Adjacent-category models for ordinal time series and their application to climate-dependent spruce budworm defoliation dynamics}

\author[a,b]{Olaloudé Judicaël Franck Osse,}
\author[a,b,c]{Zinsou Max Debaly,}
\author[b]{Philippe Marchand}
\author[a,b]{and Miguel Montoro Girona}

\affiliation[a]{Institut de Recherche sur les Forêts (IRF),
445 boul. de l'université,
Rouyn-Noranda, QC, J9X 5E4, Canada}
\affiliation[b]{Groupe de Recherche en Écologie de la MRC-Abitibi (GREMA),
341 rue principale nord, Amos, QC, J9T 2L8, Canada}
\affiliation[c]{CY Cergy Paris University, UMR CNRS 8088, 33 Bd du Port, 95000 Cergy, France}

\emailAdd{olaloudejudicaelfranck.osse@uqat.ca}

\maketitle

\vspace{0.2cm}
 
\paragraph{Abstract}
This work proposes an adjacent-category autoregressive model for time series of ordinal variables. We apply this model to dendrochronological records to study the effect of climate on the intensity of spruce budworm defoliation during outbreaks in two sites in eastern Canada. The model’s parameters are estimated using the maximum likelihood approach. We show that this estimator is consistent and asymptotically Gaussian distributed. We also propose a Portemanteau test for goodness-of-fit.  
 Our study shows that the seasonal ranges of maximum daily temperatures in the spring and summer have a significant quadratic effect on defoliation. The study reveals that for both regions, a greater range of summer daily maximum temperatures is associated with lower levels of defoliation up to a threshold estimated at 22.7 $^\circ$C (CI of 0--39.7 $^\circ$C at 95\%) in Témiscamingue and 21.8 $^\circ$C (CI of 0--54.2 $^\circ$C at 95\%) for Matawinie. For Matawinie, a greater range in spring daily maximum temperatures increased defoliation, up to a threshold of 32.5 $^\circ$C (CI of 0--80.0 $^\circ$C). We also present a statistical test to compare the autoregressive parameter values between different fits of the model, which allows us to detect changes in the defoliation dynamics between the study sites in terms of their respective autoregression structures. 

\paragraph{}

\newpage
\doublespacing
\section{Introduction}

Ordinal categorical data, i.e., data taking discrete values that can be ranked on a scale but that do not represent precise numerical intervals, are common in many  disciplines,  including  medicine, economics,  sociology and psychology, geology, wildlife studies,  geography, education, biology and ecology \citep{leonard1999course, agresti2010analysis}. They can be produced when the observed phenomenon occurs inherently on a series of discrete, qualitative states, such as the phenological stages of flowering \citep{salisbury2013flowering}. In other cases, the underlying phenomenon is continuous but is recorded on an ordinal scale to allow for the rapid surveying of a large area. Examples include regional surveys of weed density \citep{goodsell2021developing} or aerial surveys of defoliation from forest insect outbreaks (e.g., \citeauthor{mffp_tbe}). If the ordinal variable being measured is also dependent over successive points in time, then the analysis of these data requires methods adapted to ordinal time series.

Important developments have  been made recently for discrete-valued time series \citep{weiss2018introduction, davis2016handbook}. Many models exist for count data \citep{fokianos2012nonlinear, DFT}, binary data \citep{manner2016modeling,Fok1}  categorical data \citep{fokianos2003regression} and even multivariate mixed data \citep{debaly2023multivariate}. Unfortunately, ordinal categorical time series do not receive much attention. Indeed, since about 1980, authors have distinguished between ordered- and unordered-scale categorical data. 
\cite{fokianos2003regression} introduced generalized linear-type models and distinguished models for ordinal and nominal time series. Their work was extended by \cite{Fok1} for nominal data.  The latter proposed a model nested in the broad class of observation-driven \citep{cox1981statistical} models for time series data (see, for example, \cite{davis2003observation}).

In this research, we present a new observation-driven autoregressive model for ordinal categorical time series data, which we name an adjacent-category autoregressive model (hereafter \textit{ACAR}). For other methods for modeling nomimal categorical data, we refer interested readers to \cite{weiss2020regime} for advances in  so-called discrete autoregressive–moving-average model (\emph{DARMA}), presented in \cite{jacobs1978discrete}, and to \cite{Raftery} for the mixture transition distribution model (\emph{MTD}). 
The probability stability properties of the ACAR models are derived by applying the perturbation and coupling results of \cite{truquet2019coupling}. We show that the conditional maximum likelihood estimator is consistent and asymptotically Gaussian distributed. We propose Portemanteau tests \citep{akashi2021diagnostic} to verify the adequacy of the ACAR model. It is shown that under the null hypothesis of adequacy, the sum of squares of the first few autocorrelations of certain model residuals is asymptotically chi-square distribution, if scaled appropriately.  We also propose a statistical test for assessing the equality of a pair of ACAR models. Our method is then applied to tree-ring records of  spruce budworm (\textit{Choristoneura fumiferana}; SBW) defoliation in two regions of the eastern Canadian boreal forest. We investigate autoregressive effects and the exogenous effect of climate variables on outbreak development. 

\section{Materials and methods}\label{sec::material}


\paragraph{Notations and conventions}
For any sequence $(u_n)_{n\in\N},$ we adopt the convention $\sum_{i = 0}^{-1} u_i = 0$. For a positive real number $x$, we set $\log^+(x) = \log(x)$ if $x \geq 1$ and $0$ otherwise.  When $x$ is a vector of $\R^d, |x|_1$ is the standard norm-$1$ of $x$. For a $(m,n)-$matrix  $A = (a_{i,j})_{1\leq i\leq m, 1 \leq j \leq n}, \|A\|_1 = \mathrm{max}_{1\leq j\leq n} \sum_{i=1}^m |a_{i,j}|, A^\top = (a_{j,i})_{1\leq j\leq n, 1 \leq i \leq m}$ is the $(n,m)-$matrix transpose of $A$ and $A[i:j, k:\ell]$ represents the submatrix of $A$ with entries  $(a_{u,v})_{i\leq u \leq j, k \leq v \leq \ell}.$ For $A_1, \ldots, A_p$ matrices with $r$ rows and $c_1, \ldots, c_p$ columns, $A = (A_1|\cdots|A_p)$ is the column concatenation of $A_1, \ldots, A_p.$ For the two vectors $a = (a_1, \ldots, a_p)$ and $b = (b_1, \ldots, b_p)$ of $\R^p,$ $a \odot b = (a_1b_1, \ldots, a_pb_p)$ is the Hadamard product of $a$ and $b.$ For an $m\times m$ matrix $A$, the spectral radius of $A$ is denoted by $\rho(A).$

\subsection{General formulation}\label{sec::overview}

We denote $Y_t$ as the response variable. In our example in Section 3, $Y_t$ corresponds to the defoliation level for each area classified on a scale of 0 to 3, corresponding to no defoliation (approx. $0\%$) or a low (approx. $1\%-35\%$), moderate ($36\%-70\%$) or high ($71\%-100\%$) fraction of the year’s foliage defoliated by SBW. Our model is given by 
\begin{align}
\label{eq::model}
 Y_{k,t} = \mathds 1\left\{\sum_{k = 0}^{j-1} \pi_{k,t}\leq U_t < \sum_{k = 0}^{j}\pi_{k,t}\right\},  \quad & 0  \leq j \leq K, \\ 
 \label{eq::model_recursion}
 \log \frac{\pi_{j,t}}{\pi_{j-1,t}} =: \eta_{j,t} = \omega_j + \gamma^\top X_{t-1} + \alpha^\top \overline{Y}_{t-1} + \beta_j \eta_{j,t-1},  & \quad 1 \leq j \leq K,t\in\Z ,  
\end{align}
where $\omega_j\in \R, \beta_j  \in \R^*, j = 1, \ldots,K, \gamma \in \R^P\text{ and } \alpha\in \R^K$.  The vector $\overline Y_t = (Y_{1,t},\ldots, Y_{K,t}) \in \{0,1\}^K$, where for  $1\leq j \leq K, ~ Y_{j,t} = 1$ when $Y_{t} = j.$   The vector $(Y_{0,t}, \ldots, Y_{K,t})$ is then the one-hot encoding of the categorical random variable $Y_t$ valued in $\{0,\ldots,K\}.$ It is well known that the parameters $\beta_j, j= 1,\ldots, K$ act as feedback parameters; $\beta_j, j= 1,\ldots, K$  allow us to extend the memory of Model \eqref{eq::model}-\eqref{eq::model_recursion} beyond lag 1 when the effect of lagged values decreases quickly  when $j = 1, \ldots, K,|\beta_j|< 1.$

 \paragraph{Others specifications for probabilities}
 At least two other ratios can be considered in place of $\frac{\pi_{j,t}}{\pi_{j-1,t}} $ in Equation \eqref{eq::model_recursion}. For example, $\frac{\sum_{i = 1}^j \pi_{i,t}}{\sum_{i = j+1}^K \pi_{i,t}}, j= 1,\ldots, K$ corresponds to a cumulative-logit model \citep{agresti2010analysis, fokianos2003regression}, and $\frac{\pi_{j,t}}{\sum_{i = j+1}^K \pi_{i,t}}, j= 1,\ldots, K$ is a continuation-ratio logit model \citep{agresti2010analysis}. The most popular of these three ratios for an ordinal response is the cumulative-logit model. However, the adjacent-category model presents some advantages for interpretability. We detail some of these advantages below.
 
 Let us denote by $X_{p,t}$ the $p-th$ covariate and $X_{-p,t}$ the remaining covariates. 
 Setting $$R(X_{p,t}, X_{-p,t}, \overline Y_{t-1}) = \frac{\pi_{j,t}}{\pi_{j-1,t}} \text{ and } R_k(X_{p,t}, X_{-p,t}, \overline Y_{t-1}) = \frac{\pi_{j+k,t}}{\pi_{j-1,t}} ,$$ the ratio of probabilities of two successive values of $Y_t$ and this ratio when the difference between levels is $k+1$
for two different values $X_{-p,t}$ and $X_{-p,t}'$ is$$\frac{R(X_{p,t} is X_{-p,t}, \overline Y_{t-1})}{R(X_{p,t}', X_{-p,t}, \overline Y_{t-1})} = \exp\left(\gamma_p (X_{p,t} - X_{p,t}') \right) \text{ and } \frac{R_k(X_{p,t}, X_{-p,t}, \overline Y_{t-1})}{R_k(X_{p,t}', X_{-p,t}, \overline Y_{t-1})} = \left[\frac{R(X_{p,t}, X_{-p,t}, \overline Y_{t-1})}{R(X_{p,t}', X_{-p,t}, \overline Y_{t-1})}\right]^{k+1},$$ 
where $\gamma_p$ is the regression parameter of the covariate $(X_{p,t})_{t\in\Z}$. These quantities do not depend on the category $j.$

Moreover, for covariates having $\gamma_p < 0$, larger values of $X_{p,t}$  are associated with lower values of $Y_t$, and, in contrast, if $\gamma_p > 0$, as $X_{p,t}$ increases $Y_t$ is more likely to fall into higher categories.

Furthermore, Model \eqref{eq::model}-\eqref{eq::model_recursion} is similar to the category baseline logistic model for nominal data because
\begin{equation}
    \label{eq::baseline_similarity}
    \lambda_{j,t} := \log \frac{\pi_{j,t}}{\pi_{0,t}} = \sum_{k = 0}^{j-1} \log \frac{\pi_{k+1,t}}{\pi_{k,t}} = \sum_{k = 0}^{j-1} \eta_{j+1}.
\end{equation}

For $k = 1, \ldots, K$,
\begin{equation}
    \label{eq::transformation}
    \pi_{k,t} = \frac{\e^{\sum_{j = 1}^k \eta_{j,t} }}{1+\sum_{i = 1}^K \e^{\sum_{j=1}^i \eta_{j,t} }},\quad \pi_{0,t} = 1 - \sum_{k = 1}^K \pi_{k,t} .
\end{equation}

\paragraph{Equal slopes model} Equations \eqref{eq::model}-\eqref{eq::model_recursion} coincide with the equal slopes model in multinomial regression because the parameters $\alpha$ and $\gamma$ in Equation \eqref{eq::model_recursion} do not depend on $j.$ However, one can note that the coefficients of latent processes are indexed with the categories. Nonetheless, it is possible to build a full equal slopes model by replacing $\beta_j \eta_j$ in Equation \eqref{eq::model_recursion} with $\sum_{k = 1} ^ K \beta_{k} \eta_k.$ The latter model is not a nested model of the considered model and therefore cannot be compared with this model using nullity coefficients tests (see, 
 for example, Chap. 9 in \cite{heyde2013quasi}). Some arguments can be made in favor of models having specific effects, i.e., 
 \begin{align}
 \log \frac{\pi_{j,t}}{\pi_{j-1,t}} =: \eta_{j,t} = \omega_j + \gamma_j^\top X_{t-1} + \alpha_j^\top \overline{Y}_{t-1} + \beta_j \eta_{j,t-1},  & \quad 1 \leq j \leq K,t\in\Z . 
\end{align}

For the application under consideration here, our model allows us to disentangle the effect of climate on different phases of SBW outbreaks. Here, the parameter $\gamma_1$ represents the effect of covariates on the initial outbreak phase, and $\gamma_j, j> 1$ represents its outbreak development. Such an approach appears in \cite{bouchard2006forest} in which the authors apply Bayesian model averaging \citep{anderson2004model, soti2019effect}.

\paragraph{Model for more lags} 
For a more general model, Equation \eqref{eq::model_recursion} can be replaced by 
$$
\log \frac{\pi_{j,t}}{\pi_{j-1,t}} =: \eta_{j,t} = \omega_j + \gamma^\top X_{t-1} + \sum_{i = 1}^r \alpha_i ^\top \overline{Y}_{t-i} + \sum_{i = 1}^r \beta_{j,i} \eta_{j,t-i},   \quad 1 \leq j \leq K,t\in\Z
$$
for a fixed $r.$ Nonetheless, a model having one lag is often sufficient to capture the properties of a series. 
However, it is also possible to consider a long-memory model \citep{doukhan2016nonlinear},
$$
\log \frac{\pi_{j,t}}{\pi_{j-1,t}} =: \eta_{j,t} = \omega_j + \gamma^\top X_{t-1} + \sum_{i = 1}^\infty \alpha_i ^\top \overline{Y}_{t-i}    \quad 1 \leq j \leq K,t\in\Z ,
$$
although this lies beyond the scope of our work.

\paragraph{Other statistical models} Because the raw data are supposed to lie within a $(0,1)$ interval,  statistical models for $(0,1)-$distributed random variables can be used. Such models, known as beta autoregression models, are discussed by \cite{gorgi2021beta} and \cite{guolo2014beta} among others. However, in our case, some remarkable values, such as extrema $0\%$ and $100\%$ and some central values, have important empirical frequencies; therefore, models for continuous data on $(0,1)$ will not perform well in this case. 
In addition, even if the defoliation is a continuous phenomenon, small changes in its structure will not be indicated within the sample.

The following result stands for the stability properties of Equation \eqref{eq::model}-\eqref{eq::model_recursion}.

\begin{prop}
    \label{pp::solution}
    Consider the Model \eqref{eq::model}-\eqref{eq::model_recursion} and assume that $(X_t, U_t)_{t\in\Z}$ is stationary and ergodic with $\E\log_+|X_0|_1<\infty$. If for $k=1,\ldots, K, |\beta_k|<1$, then there exists a unique non-anticipative, stationary and ergotic sequence $(Y_t, X_t, \eta_t)_{t\in\Z}$ solution for the Model \eqref{eq::model}-\eqref{eq::model_recursion}. In addition, for $r\geq 1$, $\E|\eta_0|_1^r<\infty$ if $\E|X_0|_1^r<\infty.$
\end{prop}

Other more restrictive conditions for stability can be derived using the \cite{debaly2023multivariate} Theorem 1 based on \cite{debaly_truquet_2021}. Indeed, these conditions imply the parameters $\alpha$. A similar comparison can be made between the condition of Theorem 1 in \cite{Fok1} and that of Proposition 4 in \cite{truquet2019coupling}.

\subsection{Statistical inference}
From Equation \eqref{eq::model_recursion},
$$
\eta_t = \omega + \Gamma X_{t-1} + A \overline Y_{t-1} + B \eta_{t-1} ,
$$
where $\omega = (\omega_j)_{1\leq j \leq K}, \Gamma^\top = (\gamma|\cdots|\gamma),A^\top = (\alpha|\cdots|\alpha)$ and $B  = \mathrm{diag}(\beta_j, 1\leq j \leq K)$. The vector of the parameters of the model is denoted by  $\theta = (\omega^\top, \mathrm{vec}(\Gamma)^\top, \alpha^\top, \beta_1, \ldots, \beta_K)^\top$, and we assume that $\Theta (\ni \theta)$ is a compact set. The true value of the parameter is $\theta_0$. For what follows, we write $\eta_t(\theta)$ (resp. $\eta_{j,t}(\theta), \pi_{j,t}(\theta), 1\leq j \leq K)$ to make the latent process depending on the parameter $\theta.$

For $k = 1, \ldots, K$,
$$
\pi_{k,t}(\theta) = \frac{\e^{\sum_{j = 1}^k \eta_{j,t}(\theta)}}{1+\sum_{i = 1}^K \e^{\sum_{j=1}^i \eta_{j,t}(\theta) }}, \quad  \theta \in \Theta,
$$
and the parameter $\theta$ can by estimated by 
\begin{equation}
\label{eq::cmle}
    \hat \theta_n = \underset{\theta \in \theta}{\mathrm{argmin}}-\sum_{t=1}^n \left( 
    \sum_{k = 1}^K Y_{k,t}\left(\sum_{j=1}^k\eta_{j,t}(\theta)\right) - \log\left(1+\sum_{k=1}^K \e^{\sum_{j=1}^k \eta_{j,t}(\theta)}\right)
    \right)
\end{equation}
with initial values of $\eta_0$ (resp. $\overline Y_0$). This estimator corresponds to the conditional maximum likelihood estimator of the Model \eqref{eq::model}-\eqref{eq::model_recursion}. Let us set 
\begin{eqnarray*}
   s_t(\theta) = -\sum_{k=1}^K \left(\sum_{j\geq k} Y_{j,t} - \frac{\sum_{j\geq k} \e^{\sum_{i=1}^j \eta_{i,t}(\theta)}}{1+\sum_{k=1}^K \e^{\sum_{j=1}^k \eta_{j,t}(\theta)}}\right)\nabla_\theta \eta_{k,t}(\theta) \\
\end{eqnarray*}
and
\begin{eqnarray*}
  h_t(\theta)  & =  & \sum_{k=1}^K \left\{ \frac{1}{\left(1+\sum_{k=1}^K \e^{\sum_{j=1}^k \eta_{j,t}(\theta)}\right)^2}\left[\left(\sum_{j\geq k}  \e^{\sum_{i = 1}^j \eta_{i,t}(\theta)} \right) \sum_{\ell = 1}^{k-1} \left(1+\sum_{j=1}^{\ell-1}\e^{\sum_{i = 1}^j \eta_{i,t}(\theta)}\right) \nabla_\theta \eta_{\ell,t}(\theta) + \right. \right. \\
  & & \left.  \left. \left(1+\sum_{j=1}^{k-1}\e^{\sum_{i = 1}^j \eta_{i,t}(\theta)}\right) \sum_{\ell = k}^{K}\left(\sum_{j\geq \ell}\e^{\sum_{i = 1}^j \eta_{i,t}(\theta)}\right)\nabla_\theta \eta_{\ell,t}(\theta) \right]\right\} \nabla_\theta^\top \eta_{k,t}(\theta) 
\end{eqnarray*}

with $\nabla_\theta^\top \eta_{k,t}(\theta) = \left(\dpartial{\eta_{k,t}}{\theta_i}\right)_{1\leq i \leq 3K+P}$ (see Appendix for the computations).
\begin{prop}
    \label{pp::asymptotics}
    \begin{enumerate}
        \item Suppose that \begin{enumerate*}
            \item the conditions of Proposition \ref{pp::solution} hold for any $\theta \in \Theta$  with $\E|X_0|_1<\infty$ and 
            \item  $\eta_0(\theta) = \eta_0(\theta_0)~\P_{\theta_0}-a.s \Rightarrow \theta = \theta_0.$
        \end{enumerate*}  Then, the estimator (Equation \eqref{eq::cmle}) is consistent, i.e., a.s.
        $$
        \lim_{n\to\infty}\hat \theta_n = \theta.
        $$
        \item If, in addition, $\theta_0$ is located in the interior of $\Theta$, $\E|X_0|_1^2<\infty$, and the matrix $\E h_0(\theta_0)$ is invertible.
        $$
        \lim_{n\to\infty} \sqrt{n}\left(\hat \theta_n - \theta_0 \right) = \mathcal N\left(0, J^{-1}L{J^{-1}}^\top\right) ,
        $$
        where $L = \E s_0(\theta_0)s_0^\top(\theta_0)$ and $J = \E h_0(\theta_0).$
    \end{enumerate}
\end{prop}
We do not prove the previous result. Interested readers will find some elements of the proof in \cite{Fok1}, proof of Theorem 2. The same lines of proof can be found in \cite{straumann2006quasi} for the GARCH model or \cite{debaly2023multivariate} for a multivariate model for time series of mixed data. A condition for the identifiability assumption, $\eta_0(\theta) = \eta_0(\theta_0)~\P_{\theta_0}-a.s \Rightarrow \theta = \theta_0,$ based on \cite{debaly2023multivariate}, is given in the Appendix. We asume the invertibility of the matrix $J$. Finally, in the estimation procedure for an observation-driven model, one often requires the stationary condition only for the true value of parameter $\theta_0$ and the condition for contraction for the map that defines the latent process (see Chap. 7 in  \cite{francq2019garch} or \cite{straumann2006quasi}). In our case, these coincide. 

\subsection{Portemanteau-type tests for diagnostic checking}
Here, we test the adequation of the Model \eqref{eq::model}-\eqref{eq::model_recursion} and the set of assumptions of Proposition \ref{pp::asymptotics}. The null hypothesis is then that Model \eqref{eq::model}-\eqref{eq::model_recursion} holds with $\theta = \theta_0.$ One can look at the residuals:
\begin{equation}
    \label{eq::residuals}
    e_{t}(\theta) = \left(\sum_{j\geq k} Y_{j,t} - \frac{\sum_{j\geq k} \e^{\sum_{i=1}^j \eta_{i,t}(\theta)}}{1+\sum_{k=1}^K \e^{\sum_{j=1}^k \eta_{j,t}(\theta)}}\right)_{1\leq k \leq K}, t\in\Z, \theta \in \Theta
\end{equation}
and the autocorrelations:
\begin{equation}
    \label{eq::autocorrelation}
    \rho_h(\theta) = \frac{1}{n} \sum_{t=1}^n e_{t}(\theta) \odot e_{t-h}(\theta), h = 1,\ldots,n,  \theta \in \Theta .
\end{equation}
The vector of $q$ successive autocorrelations is $\rho_{1:q}(\theta) = \mathrm{vec}((\rho_1(\theta)| \cdots |\rho_q(\theta))^\top)$, and its evaluation at $\hat \theta_n$   is $\rho_{1:q}(\hat \theta_n) = \mathrm{vec}((\rho_1(\hat \theta_n)| \cdots | \rho_q(\hat \theta_n))^\top).$ The vector $\rho_{1:q}(\theta)$ evaluated at $\theta_0$ is denoted by $\rho_{1:q}.$  Note that $\rho_{1:q} = n^{-1}\sum_{t=1}^n m_t$ , and under the null hypothesis $\E(m_t|\mathcal F_{t-1}) = 0,$ i.e., $(m_t)_{t\in\Z}$, is a martingale difference sequence.
Obviously, there are many choices for residuals that make  $(e_{t}(\theta_0) \odot e_{t-h}(\theta_0))_{t\in \Z}$  a martingale difference sequence, such as  

\begin{equation}
    \label{eq::residuals_other}
   \left( Y_{k,t} - \frac{\e^{\sum_{j=1}^k \eta_{j,t}(\theta)}}{1+\sum_{k=1}^K \e^{\sum_{j=1}^k \eta_{j,t}(\theta)}}\right)_{1\leq k \leq K}, t\in\Z, \theta \in \Theta .
\end{equation} The residuals of interest here is relevant because 
$$
s_t(\theta) = -\sum_{k=1}^K e_{k,t}(\theta) \nabla_\theta \eta_{k,t}(\theta) 
$$
with $e_{k,t}(\theta)$ as the $k$-th element of $e_t(\theta)$. This allows us to derive the asymptotic distribution of $\sqrt{n}\rho_{1:q}.$

The Jacobian matrix $\xi_t$ of $e_t$ with respect to $\theta$ is 
\begin{eqnarray*}
\xi_t(\theta)(k,\cdot)^\top  & =  & - \frac{1}{\left(1+\sum_{k=1}^K \e^{\sum_{j=1}^k \eta_{j,t}(\theta)}\right)^2}\left[\left(\sum_{j\geq k}  \e^{\sum_{i = 1}^j \eta_{i,t}(\theta)} \right) \sum_{\ell = 1}^{k-1} \left(1+\sum_{j=1}^{\ell-1}\e^{\sum_{i = 1}^j \eta_{i,t}(\theta)}\right) \nabla_\theta \eta_{\ell,t}(\theta) + \right. \\
  & &   \left. \left(1+\sum_{j=1}^{k-1}\e^{\sum_{i = 1}^j \eta_{i,t}(\theta)}\right) \sum_{\ell = k}^{K}\left(\sum_{j\geq \ell}\e^{\sum_{i = 1}^j \eta_{i,t}(\theta)}\right)\nabla_\theta \eta_{\ell,t}(\theta) \right] .
\end{eqnarray*}

  For the sake of readability, we write  $\hat \rho_{1:q} $  (resp.   $e_t$  and $\xi_t$) for $\rho_{1:q}(\hat \theta_n)$ (resp.  $e_t(\theta_0)$ and $\xi_t(\theta_0)$).

Let us set $C_q = (c_1|\cdots|c_q)^\top$ with $c_k = (c_{k,1}| \cdots |c_{k,q})$. For $h = 1, \ldots, q, c_{k,h}^\top = \E e_{k,-h}\xi_0(k, \cdot), k = 1, \ldots, K$, and $\hat C_K$ is its empirical counterpart. Let us set  $$d_{i,j}({u,v}) = \E \e_{i+1,0}\e_{i+1,-u}\e_{j+1,0}\e_{j+1,-v}, i,j = 0, \dots, K-1, u,v = 1 , \dots. q,$$  $G = (g_1|\cdots | g_K)^\top$,  $g_k = (g_{k,1}^\top| \cdots | g_{k,q}^\top)$ with 
$$
g_{k,h} = \E e_{k,-h}e_{k,0}\sum_{\ell = 1}^K e_{\ell, 0}\nabla_{\theta_0}^\top\eta_{\ell,0}(\theta) {J^{-1}}^\top, 1\leq h \leq q,
$$
and $D$ is the $Kq \times Kq$ matrix with elements given by $$D[(iq+1):(i+1)q, (jq+1):(j+1)q] = [d_{i,j}(u,v)]_{1\leq u,v \leq q}, 0 \leq i,j \leq K-1,$$
where $\hat D$, $\hat G$ and  $\hat S$ are the empirical counterparts of $D$, $G$ and $S.$ Let us set $W =   D +   C_q   J^{-1}   L {  J^{-1}}^\top   C_q^\top +   G   C_q^\top +   C_q    G^\top$ and $\hat W$ as its empirical counterpart.

We require the following additional assumption, which represents the invertibility of $W.$ For $k= 1, \ldots, K$,  let us set 
$$
\iota_{k,t} = \left(0_q^\top, \ldots, 0_q^\top, \underbrace{e_{k,t-1}(\theta_0), \ldots, e_{k,t-q}(\theta_0)}_{\text{k-th block of q elements}},0_q^\top, \ldots, 0_q^\top \right)^\top.
$$

\begin{description}
    \item[INV-1\label{ass::invertible}] For $\lambda \in \R^{Kq},$ if 
    $$
    \lambda^\top \sum_{k=1}^K e_{k,0}\left(\iota_{k,0} + C_q J^{-1} \nabla_{\theta_0}\eta_{k,0}(\theta)\right) = 0 \text{ a.s.} ,
    $$
    then $\lambda = 0.$
\end{description}

\begin{prop}
    \label{pp::portemanteau}
    Under the assumptions of Proposition \ref{pp::asymptotics} with an additional moment condition on $X_0$ and \ref{ass::invertible}, as $n$ tends to $\infty,$
    $$
    n \hat \rho_{1:q} ^\top \hat W^{-1}\hat \rho_{1:q} \Rightarrow \chi_{Kq}^2 .
    $$
   Then the adequacy of the \emph{ACAR} Model \eqref{eq::model}-\eqref{eq::model_recursion} is rejected at the asymptotic level $\alpha$ if
    $$
     n \hat \rho_{1:q}^\top \hat W^{-1}   \hat \rho_{1:q} > \chi_{Kq}^2(1-\alpha) .
    $$
   
\end{prop}

It is well known that Portemanteau tests are omnibus tests \citep{francq2019garch}. They are less powerful than specific alternative tests that correspond to the nullity of the coefficient of the extra-lag model. Our study focuses only on a model of lag $1.$

\subsection{Comparison test of two adjacent-category time series models}
When we are interested in two or more sites for the same process, it may happen that we have the same models for certain sites with coefficient values that are a priori different. We propose a two-by-two model comparison test. For example, when we are interested in two ecological sites, we assume that the model on one site is driven by $\theta_0^{(1)}$ and that of the second site is driven by $\theta_0^{(2)}$. We want to test whether the $p-th$ covariate has the same effect on both sites, that is $\theta_{p,0}^{(1)} = \theta_{p,0}^{(2)}$ , and also globally $\theta_0^{(1)} = \theta_0^{(2)}$. For $j = 1,2$, let us denote $(Y_t^{(j)}, X_t^{(j)})_{t\in\Z}$ as the processes $(Y_t, X_t)$ coinciding with the  site $j$. 
The issue here is testing the equality of the distributions of $(Y_t^{(1)}, X_t^{(1)})_{t\in\Z}$ and $(Y_t^{(2)}, X_t^{(2)})_{t\in\Z}$. Tests of the laws of two strictly stationary processes have been investigated previously. They consist of tests on: \begin{enumerate*}
    \item  marginal distributions \citep{Comparingmarginaldensities, doukhan2015data};
    \item jointly finite distributions \citep{dhar2014comparison}; and 
    \item all possible $d-$dimensional joint distributions of both
processes \citep{pommeret2022testing}.
\end{enumerate*}   
With regards to the parametric form of Equation \eqref{eq::model}-\eqref{eq::model_recursion}, we can perform the test of equality of laws of entire processes without requiring their finite distributions.

For  $\theta \in \Theta$, $(s_t^{(j)}(\theta))_{t\in\Z}$ and $(h_t^{(j)}(\theta))_{t\in\Z}$ are respectively the score and Hessian processes evaluated on $\theta$ for the process $(Y_t^{(j)}, X_t^{(j)})_{t\in\Z}.$ Under the conditions of Proposition \ref{pp::asymptotics} for $(Y_t^{(j)}, X_t^{(j)})_{t\in\Z},$
$$
\lim_{n\to\infty} \frac{1}{\sqrt{n}}\sum_{t=1}^n s_t^{(j)}(\theta_0^{(j)})= \mathcal N(0,L_j) 
$$
and \emph{a.s}.
$$
\lim_{n\to\infty} \frac{1}{n}\sum_{t=1}^n h_t^{(j)}(\theta_0^{(j)}) = J_j, j=1,2 .
$$
Let us set $S =\E s_0^{(2)}(\theta_0^{(2)}) s_0^{(1)}(\theta_0^{(1)})^\top$, $V = J_1^{-1}L_1{J_1^{-1}}^\top + J_2^{-1}L_2{J_2^{-1}}^\top + J_2^{-1}  S {J_1^{-1}}^\top  + {J_1^{-1}} S^\top  {J_2^{-1}}^\top$ and $\hat V$ its empirical counterpart. We denote $\hat \theta_n^{(1)}$ (resp. $\hat \theta_n^{(2)}$) the estimator of $\theta_0^{(1)}$ (resp. $\theta_0^{(2)}$), given by Equation \eqref{eq::cmle} , and $\hat \theta_{p,n}^{(1)}$ (resp. $\hat \theta_{p,n}^{(2)}$) as the $p-th$ element of $\theta_0^{(1)}$ (resp $\theta_0^{(2)}$).

We require the following additional assumption for the invertibility of $V$.
\begin{description}
    \item[INV-2\label{ass::invertible2}] For  $\lambda \in \R^{3K+P},$ if 
    $$
    \lambda^\top  \left( J_1^{-1}s_t^{(1)}(\theta_0^{(1)}) +  J_1^{-2}s_t^{(2)}(\theta_0^{(2)}) \right)= 0 \text{ a.s.},
    $$
    then $\lambda = 0.$
\end{description}
\begin{prop}
\label{pp:comparison}
Suppose that  $\left(s_t^{(1)}(\theta_0^{(1)}),s_t^{(2)}(\theta_0^{(2)})\right)_{t\in\Z}$ is stationary and ergodic. Under the assumptions of Proposition \ref{pp::asymptotics} and  \ref{ass::invertible2}:
 \begin{enumerate}
   \item As $n$ tends to $\infty,$
     $$
     n^{1/2} \frac{\hat \theta_{p,n}^{(1)} - \hat \theta_{p,n}^{(2)}}{\sqrt{\hat V(p,p)}}  \Rightarrow  \mathcal N(0,1).
     $$
    The hypothesis $\theta_{p,0}^{(1)} = \theta_{p,0}^{(2)}$ is then rejected at the level $\alpha$ if 
    $$ 
    \left|n^{1/2} \frac{\hat \theta_{p,n}^{(1)} - \hat \theta_{p,n}^{(2)}}{\sqrt{\hat V(p,p)}}\right| > q(1-\alpha/2).$$
    
     \item As $n$ tends to $\infty,$
     $$
     n \left(\hat \theta_n^{(1)}- \hat \theta_n^{(2)}\right)^\top \hat V^{-1} \left(\hat\theta_n^{(1)}- \hat\theta_n^{(2)}\right) \Rightarrow \chi^2_P.
     $$
    The hypothesis $\theta_0^{(1)} = \theta_0^{(2)}$ is then rejected at the level $\alpha$ if 
    $$ 
    n \left(\hat \theta_n^{(1)}- \hat \theta_n^{(2)}\right)^\top \hat V^{-1} \left(\hat\theta_n^{(1)}- \hat\theta_n^{(2)}\right) > \chi^2_P(1-\alpha).$$
 \end{enumerate}
\end{prop}

\paragraph{On the stationary assumption for scores} Note that we require the joint process of scores $\left(s_t^{(1)}(\theta_0^{(1)}),s_t^{(2)}(\theta_0^{(2)})\right)_{t\in\Z}$ to be stationary and ergodic. This assumption is satisfied as soon as the process $\left(X_t^{(1)}, X_t^{(2)}, U_t^{(1)}, U_t^{(2)}\right)_{t\in\Z}$ is stationary and ergodic, where ${(U_t^{(1)})}_{t\in\Z}$ (resp. ${(U_t^{(2)})}_{t\in\Z}$) is the process of uniform variable that generates $(Y_t^{(1)})_{t\in\Z}$ (resp. $(Y_t^{(2)})_{t\in\Z}$. The case for the independency of the processes $\left(X_t^{(1)},  U_t^{(1)} \right)_{t\in\Z}$ and $\left(X_t^{(2)}, U_t^{(2)}\right)_{t\in\Z}$ is particularly interesting because it entails S = 0. It can be also interesting to consider the case where $\left(X_t^{(1)}\right)_{t\in\Z}$ and $\left(X_t^{(2)}\right)_{t\in\Z}$ are independent, but $\left(U_0^{(1)},  U_0^{(2)}\right)_{t\in\Z}$ is degenerate; for example $U_0^{(1)} = U_0^{(2)}$ or $U_0^{(1)} = 1 - U_0^{(2)}.$ We will investigate the latter examples in the numerical experimentation section.

\subsection{Model implementation}
The numerical implementation of Model \eqref{eq::model}-\eqref{eq::model_recursion} and the statistical inference methods are implemented using \emph{R} \citep{r_language}. We choose $\beta_{j,0} = 0.5, j = 1, \ldots, K$ for the initial values of the latent processes, whereas the first observation of the paths acts to initialize the sequence $(Y_t)_{t\in\Z}$. The initial value of the score sequence is chosen to be equal to $0_{3K+P}.$  The loss function (Equation \eqref{eq::cmle}) is optimized for $20$ different random initializations using the \emph{L-BFGS-B} method \citep{byrd1995limited},  a derivative-free optimization routine that allows box constraints, run in the R function \emph{optim}. The estimated $\theta$ values are those corresponding to the minimum loss function value over all initializations. Let us set $\theta_j$ as the $j-th$ element of the parameters vector $\theta$, and the set $\Theta \ni \theta$ is
{\footnotesize $$
\Theta =: \Theta_\varepsilon =  \left\{\theta \in \R^{3K+P}: \theta_{3K+P-j} \in [-1+\varepsilon, 1-\varepsilon], j = 1,2, 3 \text{ and } \theta_j \in [-1/\varepsilon,1/\varepsilon], j = 1, \ldots, 2K+P\right\} 
$$}
for a fixed and known $\varepsilon > 0$. Here we fix $\varepsilon = 1\e^{-6}$. All methods are made available using two top level functions \emph{poly\_autoregression} and \emph{comparison\_test\_dependent}. The first takes as an argument the observed paths of sequences $(\overline Y_t)_{t \in \Z}$ and $( X_t)_{t \in \Z}$ and returns, among others, the estimated values of parameters $\theta$, its standard values and the result of the Portemanteau test in Proposition \ref{pp::portemanteau}. The second takes as arguments the returned values of two models fitted with the \emph{poly\_autoregression} function and gives the statistics and the \textit{p}-values of the test in Proposition \ref{pp:comparison}.

\subsection{Numerical experimentation} 
Recall that the vector of the parameters is denoted by 
$\theta = (\omega^\top, \mathrm{vec}(\Gamma)^\top, \alpha^\top, \beta_1, \ldots, \beta_K)^\top$. For numerical experimentation, we use a simulated adjacent categorical response variable with four levels, as for our real data example, and five simulated covariates.  We consider the sets of parameters of Table \ref{tab::paramaters_table},  chosen to meet the condition $|\beta_j| < 1, j = 1, \ldots K$ in Proposition \ref{pp::solution}.

\begin{table}[!h]
    \centering
    \begin{tabular}{r|llllllllllllll|}
    \cline{2-15}
   \multicolumn{1}{c|}{} & $\omega_1$ & $\omega_2$ & $\omega_3$ & $\gamma_1$ &  $\gamma_2$ &  $\gamma_3$ &  $\gamma_4$ &  $\gamma_5$ & $\alpha_1$ & $\alpha_2$ & $\alpha_3$ & $\beta_1$ & $\beta_2$ & $\beta_3$\\ \hline
      \rowcolor[gray]{.9}$\theta_0^{(1)}$   & 1.2 &  0.7 &  0.5 &  -0.8 &  1.5 &   -1.5 &  2.0 &  2.0 &  0.3 &  -0.3 & 0.5 &  0.8 &  -0.2  &  0.3 \\
       $\theta_0^{(2)}$    &  1.2 &  0.7 &  0.5 &  0.8 &  -1.5 &   1.5 &  -2.0 &  -2.0 &   -0.3 &   0.3 &   -0.5 &  -0.8 &  0.2  &   -0.3 \\
       \rowcolor[gray]{.9}$\theta_0^{(3)}$ & 1.2 &   0.7 &   1.5 &   0.8 &  -1.5 &  -1.5 &   2.0 &  -2.0 &   0.3 &  -0.3 &  -0.5 &  -0.8 &   0.2 & -0.3 \\ \hline
    \end{tabular}
    \caption{The sets of parameters used in the numerical experimentation}
    \label{tab::paramaters_table}
\end{table}

\paragraph{Finite sample performance of the maximum likelihood estimator}

We investigate the finite sample performance of the estimators (Equation\eqref{eq::cmle}) for the set of the parameters $\theta_0^{(1)}, \theta_0^{(2)}$  and $\theta_0^{(3)}$.
For each value of $\theta_0,$ the data generation process consists of drawing the covariates $(X_t)_{t=1..n}$ from a standard Gaussian random distribution on $\R^5$  and $(Y_t)_{t=1..n}$, according to Equation \eqref{eq::model}. We consider four sample sizes: $n = 70, 100, 300 \text{ and } 500.$ We repeat this process $B=499$ times, and for each sample we compute the estimate $\hat \theta_n$ and the covariance matrix of Estimator \ref{pp::asymptotics}. Table \ref{tab::simulations} in the Appendix presents the simulation results. The line CMLE refers to the average  estimated values for the parameters, and TSE refers to the mean  of the estimated values for  theoretical standard error. 
$$
\mathrm{CMLE} = B^{-1} \sum_{b = 1}^{B} \hat \theta_T^{(b)} \text{ and } \mathrm{TSE} = B^{-1} \sum_{b = 1}^{B} \mathrm{diag}\left\{ {\hat J^{-1(b)}}  {\hat L}^{(b)} {\hat J^{-\top(b)}}\right\}^{1/2} ,
$$
where the superscript $b$ represents the index of replication and $\mathrm{diag}M$ for a matrix $M$ is the diagonal elements of $M$. MAE and MSE refer to the mean absolute and mean squared errors of the estimator.  It appears that the model parameters are well estimated even for $n=100.$  The estimates are less accurate for $n=70.$ Nevertheless, the true values of parameters are recovered in many cases by the confidence interval at a  $0.95$ level.

\paragraph{Comparison tests}
We set four scenarios to access the quality of Comparison test \ref{pp:comparison}. 
\begin{enumerate}
    \item The processes $\left(X_t^{(1)},  U_t^{(1)} \right)_{t\in\Z}$ and $\left(X_t^{(2)}, U_t^{(2)}\right)_{t\in\Z}$ are independent, and the true values of the parameters are both equal to $\theta_0^{(1)}$.
    \item The processes $\left(X_t^{(1)},  U_t^{(1)} \right)_{t\in\Z}$ and $\left(X_t^{(2)}, U_t^{(2)}\right)_{t\in\Z}$ are independent, and the true values of the parameters are $\theta_0^{(1)}$ and $\theta_0^{(3)}$.
    \item The processes $\left(X_t^{(1)} \right)_{t\in\Z}$ and $\left(X_t^{(2)}\right)_{t\in\Z}$ are independent, $U_0^{(2)} = 1 - U_0^{(1)}$, and the true values of the parameters are both equal to $\theta_0^{(1)}$.
     \item The processes $\left(X_t^{(1)} \right)_{t\in\Z}$ and $\left(X_t^{(2)}\right)_{t\in\Z}$ are independent, $U_0^{(2)} =  U_0^{(1)}$ and the true values of the parameters are $\theta_0^{(1)}$ and $\theta_0^{(3)}$. 
\end{enumerate}
For each of theses scenarios, the length of the paths is $n = 500$, and we perform $B = 1000$ times the comparison tests. For scenarios 2 and 4,  the null hypothesis is rejected for all $B = 1000$ tests, showing a test power near 1. For Scenario 1, the null hypothesis is accepted in 65\% cases, and this rate drops to 60\% for Scenario 3.

\section{Application to spruce budworm defoliation data}

Natural and anthropogenic disturbance regimes are majors drivers of change in the structure and function of forest ecosystems \citep{deGranpre2018incorporating, aakala2023millennial}. SBW is the most important defoliator of conifer trees in North America \citep{lavoie2019vulnerability, Efficacy_alvaro2022, kayes2020boreal, anderegg2015tree}. This defoliation and the subsequent tree mortality result in marked losses in forest productivity \citep{grondin1996ecologie} and reduce the economic and ecosystem and services provided by these trees and forests  \citep{subedi2023climatic, scott2023salvage, sato2023dynamically, chagnon2022impacts}. These defoliation events also reduce the boreal forest's carbon sequestration capacity \citep{liu2019simulation, liu2020aerial}. Damage occurs when SBW larvae repeatedly feed on the annual foliage of mature balsam fir (\textit{Abies balsamea}), white spruce (\textit{Picea glauca}) and black spruce (\textit{Picea mariana}). This defoliation reduces production, leading to radial growth suppression and eventual tree mortality \citep{lavoie2021does, liu2022evaluating, debaly2022autoregressive}.  Reduced tree-ring growth can therefore be an indirect indicator of defoliation events. Moreover, defoliated trees are more susceptible to fungus and disease, which leads to a reduction in the quality of the wood available for timber-based industries \citep{maclean2016impacts, safranyik2010potential, chang2012economic, Efficacy_alvaro2022}.

The spread of forest pest epidemics is a complex phenomenon that can occur across a wide spatial scale \citep{bouchard2014influence, montoro2018secret}. 
In eastern Canada, the boreal forest is an integral part of the environment, economy, culture and history \citep{kayes2020boreal}. Given the ecological and economic consequences of SBW outbreaks, SBW-related defoliation is a major issue in Canadian forestry.  

The reconstruction of the dynamics of SBW epidemics over the last centuries shows that the frequency, severity and spatial distribution of SBW outbreaks have changed \citep{simard2006millennial, navarro2018changes, berguet2021spatiotemporal, girona2023challenges}. SBW outbreaks are shifting in a north-eastern manner in boreal Canada \citep{pureswaran2018forest, jactel2019responses,navarro2018changes, seidl2017forest, regniere2019influence}. These changes can be explained by changes in the spatial distribution of SBW hosts because of climate shifts and climate effects on SBW population dynamics \citep{regniere2019influence, d2023building, maclean2016impacts, schowalter2022insect}. Accurate prediction of SBW spread requires understanding the interactions between SBW defoliation and climate \citep{pureswaran2015climate, overpeck1990climate, regniere2019influence}.

Understanding the interactions of climate with natural disturbances is crucial for adapting forest sustainable management to climate change \citep{seidl2023modeling, gauthier2023ecosystem}. SBW is also integrated within complex food webs \citep{eveleigh2007fluctuations}. Interactions with higher trophic levels (predators, pathogens and parasitoids) and lower trophic levels (trees) play a major role in controlling SBW population dynamics. Climate change will create extreme conditions that negatively affect the physiological mechanisms of SBW host species (photosynthesis, evapotranspiration, etc.) and the development, reproduction and mortality rates of SBW parasites and parasitoids \citep{bouchard2014influence, Candau_forcasting, jactel2019responses, li2020previous}. It should be noted that climate change will not only affect the vulnerability of the forest to defoliation but also the SBW population dynamics responsible for this defoliation \citep{pureswaran2018forest, jactel2019responses, boakye2022insect}. New methodological approaches for understanding climate and natural disturbance interactions are thus essential to develop reliable forecasts to guide forest management strategies in the face of climate change \citep{gauthier2023ecosystem}.

\subsection{Existing models of forest insect outbreaks}
Assessing the effects of climate on forest insect outbreaks is an ongoing research topic \citep{Candau_forcasting, navarro2018changes,  rs10030360, EICKENSCHEIDT20192932012, berguet2021spatiotemporal}. Using the dendrochronological series of growth rings of spruce, \cite{navarro2018changes} and \cite{berguet2021spatiotemporal} reconstructed the spatiotemporal dynamics of 20th-century SBW outbreaks across the insect's range in Quebec in forests across several regions. They then enriched their analysis by including the climate characteristics of each forest. \cite{berguet2021spatiotemporal} found three patterns in the time series of defoliation. They applied  k-means clustering \citep{hastie2009elements, bishop2006pattern, shalev2014understanding} to group the annual percentage of trees affected by SBW. One of these clusters was associated with lower annual temperatures; the other clusters corresponded to higher amounts of precipitation.  The  attempt of \cite{navarro2018changes} to distinguish patterns of severity and frequency of SBW distribution paths was not very fruitful. The authors mentioned that the statistical method consisting of ordinary least squares linear regression \citep{hastie2009elements, bishop2006pattern, shalev2014understanding} of percentage of affected trees against climate data employed could be improved by other approaches. 

\cite{Candau_forcasting} investigated the response of SBW defoliation in neighboring Ontario to climate change from a predictive point of view. They used random forest regression  \citep{hastie2009elements, bishop2006pattern, shalev2014understanding} to relate the frequency of moderate-to-severe defoliation to climate conditions. This work measured the predictive strength of each variable via the decrease in prediction accuracy (i.e., the difference in MSE) between the out-of-bag samples and the out-of-bag samples after a random permutation with respect to the variable. They determined that seasonal temperatures in winter and minimum temperatures in summer and spring are relevant when forecasting the frequency of defoliation. 

\cite{EICKENSCHEIDT20192932012} used GAM models \citep{hastie2017generalized} and collected defoliation data to associate climate factors to defoliation events in Germany between 1989 and 2015. They  reported a significant influence of climatic factors (drought stress, the strongest water deficits, etc.) and a drought-related  increase in defoliation, relying on the studies of \cite{FERRETTI201456, ZIERL200425, Kamman, seidling2007signals}. These earlier papers demonstrated that lagged effects, especially drought in the preceding year and cumulative drought over several preceding years, have a major influence on defoliation in the following year.

\cite{rs10030360} used Landsat data to analyze the defoliation of pine forests (\textit{Pinus tabulaeformis}) around Beijing. They applied ordinary least squares linear regression \citep{hastie2009elements, bishop2006pattern, shalev2014understanding} to the remaining foliage against  rainfall in the previous October and total  annual hours of sunshine. They determined that \begin{enumerate*}
    \item high active accumulated temperatures suppress the occurrence of the caterpillar and favor the recovery of trees;
    \item heavy rainfall also suppresses outbreaks, especially rainfall in the preceding October; and
    \item extensive sunshine can promote outbreaks and activate the insect. 
   The influence of these climate variables on caterpillar abundance would be expected given the biological characteristics of this particular caterpillar, confirming the reliability of their predictive model.
\end{enumerate*}

\paragraph{Our contribution}
We contribute to this debate by applying the ACAR model to  defoliation data from eastern Canadian forests. We focus our efforts on methods for interpreting data in contrast with empirical forecasting methods such as random forest \citep{Candau_forcasting}. 
Defoliation data are typically analyzed using predictive methods such as random forest \citep{Candau_forcasting},  parametric statistical models like linear regression \citep{rs10030360, germain2021insectivorous} and semi-parametric methods such as GAM models \citep{EICKENSCHEIDT20192932012, perrier2021budworm,boakye2022insect}. However, all these methods do not account for the defoliation level being an ordinal time series or the process of defoliation being cumulative over time \citep{regniere2007ecological, chen_low_level, sturtevant2015modeling}. 

Rather than applying the usual method for repeated measures, which consists of fitting a global model  for all series with as many parameters as possible being constant across series, we fit one model per series. Thus, we can highlight the dissimilarities between the defoliation processes for each region and the specific effects of covariates. From a practical point of view, the novelty of our methodology lies in its region-centric modeling. Our study therefore represents a beneficial region-adapted tool for intervention policy in forest resource management. 

\subsection{Study area}

The study area is located in the northern temperate zone of northeastern North America in the temperate forests of the province of Québec, Canada. We selected the two study sites (Témiscamingue and Matawinie) on the basis of the availability of long time series of dendrochronological and climate data.

\begin{figure}[!ht]
    \centering
    \includegraphics[scale = 0.83]{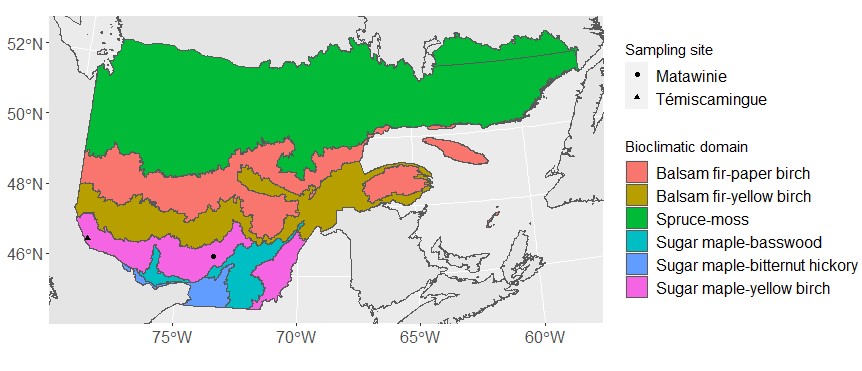}
    \caption{Location of study sites in southern Québec and overview of the boreal ecoregions.
}
    \label{fig:area}
\end{figure}

The Abitibi-Témiscamingue region is situated in western Quebec, and the regional climate  is cold and dry (Dfc, K\"{o}ppen--Geiger classiﬁcation) \citep{kottek2006world} with an average annual rainfall of approximately 900 mm (\emph{MDDEP, 2012}). 
The sample site is at 46$^{\circ}34^\prime$$ \text{ N and }78$$^{\circ}50^\prime$ W within the northern temperate zone, comprising deciduous and mixed deciduous and coniferous forest, as part of the sugar maple--yellow birch bioclimatic zone (Figure \ref{fig:area}).

Matawinie lies the administrative region Lanaudière in south-central Québec. Regional climate is cold (Dfb, K\"{o}ppen--Geiger classiﬁcation) \citep{kottek2006world} and mean annual rainfall is about 1000 mm (\emph{MDDEP, 2012}). The sampled forest is at 46$^{\circ}26^\prime$ N and 73$^{\circ}30^\prime$ W. Similar to the sampled forest in Témiscamingue, this eastern study site also lies in the northern temperate zone, comprising deciduous and mixed forests, and the sugar maple--yellow birch bioclimatic zone (Figure \ref{fig:area})

\subsection{Data description }

\subsubsection{Defoliation data}
For both study sites, we obtained 75-year (1920--1995) records of dendrochronological data from \cite{navarro2018changes}. The time period selected for this study ensures that tree-ring samples record multiple 20th-century insect outbreaks. This data set incorporates sites used in \cite{jardon2003}. In this latter survey, the authors sampled 400 m$^2$ plots and collected wood disks from seven dominant living trees and three dominant living saplings. They used dendrochronology to reconstruct SBW outbreaks in 126 locations within the eastern boreal forest and studied a large sample of trees at different forest sites, using reduced growth is an indirect indicator of defoliation. Thus,   \cite{jardon2003} calculated the average radial growth of the trees in the sample and inferred different levels of defoliation in the forest sites as a function of growth lag. We refer interested readers to \cite{navarro2018changes} for details regarding the dendroecological  methodology.

We then estimated the proportion of trees affected by defoliation at each site over the 75 years. We considered a tree to be defoliated when we observed a growth reduction for the sample that was more than 1.28 SD from the mean standardized dendrochronological record. Outbreak severity was defined as the proportion of trees at each site showing a pattern of growth reduction at a given time. We defined defoliation severity as: Level $0$ (null defoliation) corresponded to no trees showing a growth reduction greater than 1.28 SD;
     $1$ (light defoliation) reflected 1\%--$35 \%$ of trees experiencing a growth reduction; 
    $2$ (moderate defoliation) represented 36\%--$70 \%$ of trees with reduced growth; and
    $3$ (severe defoliation) reflected 71\%--$100\%$ of trees showing reduced growth.
 The data sets are available on \url{https://figshare.com/articles/dataset/Untitled_Item/7887746}.

\subsubsection{Climate data sets}

Climate data were obtained from the National Centers for Environmental Information \url{https://www.ncei.noaa.gov/}. The weather station in Témiscamingue is located 25 km away from the forest site and the distance between the Matawinie site and the nearest weather station is about 35 km. Both data sets include daily maximum and minimum temperatures, precipitation and snow depth from 1920 to 1995 (75 years of daily data). Daily climate data for 1928 contains much missing data; we therefore excluded 1928 from our analysis. From this raw data, we computed the seasonal covariates for summer and spring given the biological characteristics of SBW and following multiple earlier studies \citep{debaly2022autoregressive,volney2007spruce, volney2000climate}. We considered April, May and June as being spring and  July, August and September as summer. The specific covariates used are described in the following section. 
Given that the effect of defoliation on growth is delayed, our model includes the effect of climate covariates with a lag of two years.

\subsection{Data analysis}
From a preliminary descriptive analysis that consists of calculating the serial correlations, we considered two sets of climate covariates. We used the interannual difference (current year minus the previous year) of the mean daily maximum and mean daily minimum temperatures for the spring and summer, the seasonal range (maximum minus minimum across all days) of daily minimum and daily maximum temperatures for spring and summer, and the interannual difference in the log of annual precipitation, the log of annual snowfall, the log of annual precipitation and log of annual snowfall. For all temperature-related covariates, we applied a quadratic term to the model as well as a linear effect.  
For both sites, we fitted the Model \eqref{eq::model}-\eqref{eq::model_recursion} with all possible combinations  of our covariates. For each fitted model, we performed a Portemanteau test (see Proposition \ref{pp::portemanteau}) with one lag. Only models for which H$_0$ was not rejected were considered.  We then performed a global comparison of the retained models with the test (Proposition \ref{pp:comparison} 2.). When the null hypothesis was rejected, we carefully determined the variable that drove the deviation from H$_0$ by using  Proposition \ref{pp:comparison} 1. When the Portemanteau test did not reject the adequacy of the models of interest, we selected the best model on the data of the same forest using the nonformal criterion of the number of significance covariates at a 95\% level because both models had the same number of covariates. We compared nested models for the same forest with the \emph{AIC} information criterion. All models with an estimated $\beta_j, j = 1,\ldots, K$ equal to the upper or lower bounds (1-$\epsilon$  or -1+$\epsilon$) were not included in our analysis.

\section{Results}\label{sec::results}
The best models for the Témiscamingue and Matawinie sites were those having the seasonal range of daily maximum and minimum temperatures as covariates (see Table \ref{tab::best_selected_models}). The best model selected for our two sites suggested not only that the effect of temperature on defoliation dynamics was nonlinear but that this effect also varied according to season (summer and spring). The estimated coefficients for summer temperature and spring temperature showed a quadratic effect of temperature with a maximum and minimum, respectively, for summer and spring temperatures. Both regions showed that up to a certain temperature threshold (see details in the Appendix)---estimated at 22.7 $^\circ$C (CI of 0--39.7 $^\circ$C at 95\%) for Témiscamingue and 21.8 $^\circ$C (CI of 0--54.2 $^\circ$C at 95\%) for Matawinie---a greater range of the summer daily maximum temperatures was associated with lower levels of defoliation. Over the entire period covered by our study, summer daily maximum temperatures used for modeling ranged between 15 $^\circ$C and 30 $^\circ$C, with a median of 23.3 $^\circ$C for Témiscamingue. These values were between 15 $^\circ$C and 28.3 $^\circ$C, with a median of 21.1 $^\circ$C for Matawinie.

The best model for Matawinie showed that a greater range in spring temperature was associated with a higher level of defoliation. We estimated a threshold of 32.5 $^\circ$C (CI of 0--80.0 $^\circ$C) under which an increase in temperature promoted defoliation. Spring daily maximum temperatures in Matawinie ranged between 25 and 37.8 $^\circ$C, with a median of 31.1 $^\circ$C. Indeed, warmer springs are characterized by the premature emergence of larvae from dormancy \citep{volney2000climate, volney2007spruce}.

The comparison of these regions revealed that defoliation processes differed between the sites. This difference related to their autoregressive structure. The model for Témiscamingue was similar to a parameter-driven model (i.e., the latent process does not depend on previous values of the observed process), as the \textit{p}-value of the test H$_0 : \alpha_1 = \alpha_2 = \alpha_3 = 0$ was $0.512$. This was not the case for Matawinie.  We also denoted the statistical significance of the feedback parameters, in particular  $\beta_2$ for both regions, indicating a link between $Y_t$ and its history. This result is consistent with the known cumulative effects of defoliation on tree growth \citep{chen_low_level, houndode2021predicting}. 


\begin{table}[!ht]
\footnotesize
    \caption{Estimate of selected best models for Témiscamingue and Matawinie}
    \label{tab::best_selected_models}
\begin{tabular}{lrrrc}
& & & & \\
\cline{2-5}
&\multicolumn{4}{c}{\emph{Number of parameters} = 14    { } {} { } {}     \emph{n  } = 72 } \\
\cline{2-5}

& Témiscamingue & Matawinie & \multicolumn{2}{c}{Comparison of both models}  \\ \cline{2-5}
& Parameters & Parameters & \textit{t}-statistics & Signif. code\\
& (t-stat)  & (t-stat)  &  &  \\ 
\hline
\rowcolor[gray]{.9}$\omega_1$ & -0.684 & 17.808 & 0.144 &    \\ 
\rowcolor[gray]{.9} & (-0.005)  & (0.635)  &  &   \\  
$\omega_2$ & -35.239 & -2.233 & 0.149 &    \\ 
 & (-0.160)  & (-0.084)  & \\  
\rowcolor[gray]{.9}$\omega_3$ & -35.745 & -6.407 & 0.133 &   \\   
\rowcolor[gray]{.9} & (-0.163)  & (-0.244) &  &   \\  
$\Delta$\emph{temp. max. spring} & 56.351 & 31.206\textbf{*}  & 0.254 &   \\  
 & (0.564)  & (2.299)  &   \\ 
\rowcolor[gray]{.9}$\Delta$\emph{temp. min. summer}  & -2.454 & 1.248 & 0.543 &   \\ 
\rowcolor[gray]{.9} & (-0.366) & (1.061)  &  & \\  
$\Delta$\emph{temp. max. summer} & -58.714\textbf{***} & -67.179\textbf{*} & 0.253 &    \\ 
 & (-3.820) & (-2.198)  & \\  
\rowcolor[gray]{.9}\emph{Square of }$\Delta$\emph{temp. max. spring} & -10.742 & -4.794\textbf{*} & 0.317 &  \\ 
 \rowcolor[gray]{.9}  & (-0.568) & (-2.203)  &  & \\  
 \emph{Square of }$\Delta$\emph{temp. max. summer} &  12.927\textbf{***} & 15.404\textbf{*}& 0.338 &   \\  
  & (4.966) & (2.219) & \\  
 \rowcolor[gray]{.9}$\alpha_1$ & 34.638 & 3.201\textbf{*} & 0.310 &   \\
 \rowcolor[gray]{.9} & (0.341) & (2.004)  & &  \\  
$\alpha_2$ & 40.594 & 25.209\textbf{***} & 0.136 &    \\  
 & (0.357) & (3.264) & \\  
\rowcolor[gray]{.9}$\alpha_3$ & 42.134 & 27.066\textbf{***} & 0.132 &   \\ 
\rowcolor[gray]{.9} & (0.367)  & (3.111) &  & \\  
$\beta_1$ & 0.839 & 0.263 & 0.824 &     \\   
 & (1.348)  & (0.669) & \\  
\rowcolor[gray]{.9}$\beta_2$ & 0.811\textbf{***} & -0.985\textbf{***} & 9.288 & \textbf{***} \\  
\rowcolor[gray]{.9} & (4.212) & (-52.428) &  & \\  
$\beta_3$ & 0.693\textbf{***} & -0.009 & 0.460 &    \\
 & (3.614) & (-0.005) & \\  
\hline
AIC : &48.162 &90.400 & \multicolumn{2}{l}{Comparison stat : 12006.0***} \\
Portemanteau stat : &2.238  &0.082  &    \multicolumn{2}{l}{Comparison \textit{p}-value : <  $ 10^{-5} $ } \\ 
Portemanteau \textit{p}-value : & 0.524&0.994 &  & \\ \hline 
\multicolumn{5}{l}{\scriptsize{$\Delta$temp. : range of temperature over season }}  \\
\multicolumn{5}{l}{\scriptsize{Signif. code :   0.90 . 0.95 * 0.99 ** 0.999 *** 1}}  
\end{tabular}
\end{table} 

\section{Discussion and perspectives } \label{sec::conclusion}

Ecological studies that involve time series are generally based on intensive studies of restricted populations or small-scale ecosystem recovery. Ecological variability often occurs in a few places and under relatively static conditions \citep{goodsell2021developing}. It is therefore advantageous to cover a larger area more quickly, simply by estimating the data on an ordinal scale. Here, we propose a statistical tool for modeling adjacent-categories time series. Using the intensity of defoliation data as a case study, we relate the impact of climate variation on intensity of SBW defoliation in eastern Canadian forests through a categorical-adjacent autoregressive model. Our proposed model provides a useful tool for a more in-depth analysis of the impact of climate on defoliation because each successive defoliation level can be influenced differently by environmental factors during a SBW outbreak \citep{bouchard2014influence, jepsen2009phase}, and SBW development also depends on defoliation in previous years \citep{bouchard2014influence}.

\begin{figure}[!ht]
    \centering
    \includegraphics[scale = 0.18]{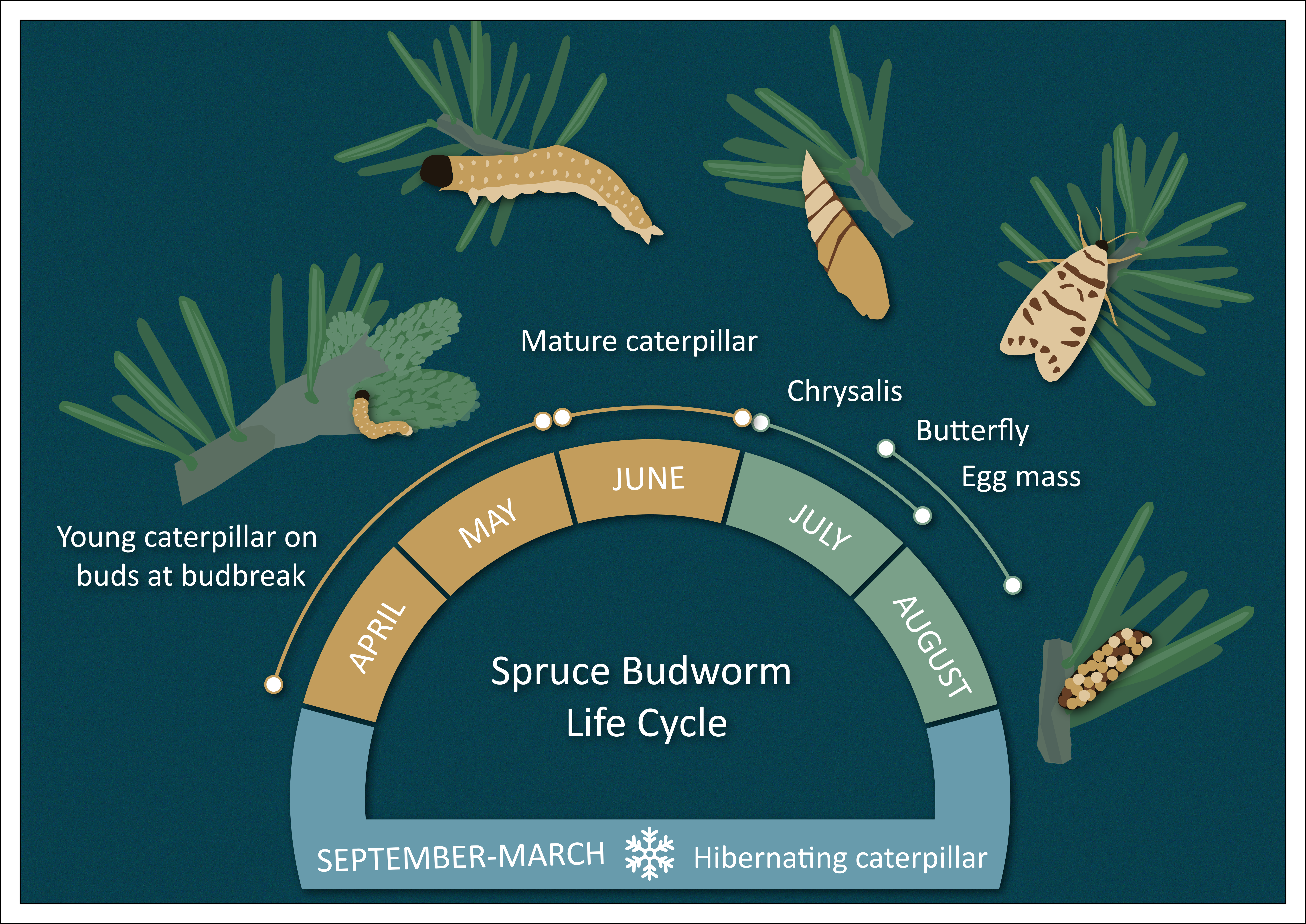}
    \caption{The life cycle of spruce budworm }
    \label{fig:SBW}
\end{figure}

Our paper presents two majors findings. 
Although several studies show that increased temperatures (caused by climate change) increase in the frequency and severity of SBW outbreaks \citep{overpeck1990climate,regniere2007ecological, pureswaran2015climate, bouchard2014influence, bouchard2018bottom, moise2019density, pureswaran2019phenological}, we observe the same effects of warmer summer temperature; however, the effect from a threshold was reversed. In fact, extreme temperatures reduce SBW fecundity, increase mortality and affect the sex ratio via an impact on endosymbionts by favoring the production of male individuals \citep{hance2007impact}. Moreover, certain temperatures favor the population dynamics of the SBW, but when this temperature preference range is exceeded, this factor negatively affects SBW. Summer temperature has a much greater influence on SBW population recruitment. Summer marks the development of larvae and the reproduction of adult SBW, life cycle steps that are closely linked to a number of climatic factors, including temperature \citep{bouchard2014influence, pureswaran2015climate, volney2000climate,pureswaran2019phenological}. Indeed, summer corresponds to the period when female  SBW deposit their eggs (see Figure \ref{fig:SBW}). Our results suggest that high variations in temperature induce a low survival egg rate. SBW density growth is determined by the egg recruitment rate and generation survival rate, e.g., \cite{royama2005analysis}, because the success of SBW in establishing itself on sites depends on reproduction, the initial weight of the eggs and the synchronization of their development with that of the buds of their hosts, which are all strongly influenced by climatic factors \citep{volney2000climate, volney2007spruce}.

We also denote a negative effect of warmer annual spring temperatures on defoliation levels. Spring is the critical period for defoliation, as it marks the emergence of SBW larvae from hibernation and the start of needle mining by SBW hosts. Early spring is generally marked by mild temperatures, which favors cohabitation between SBW and its parasitoids \citep{bouchard2018bottom, royama2005analysis}. This cohabitation could explain the negative effect of temperature in spring. SBW benefits from the negative synergistic effects of temperature pressures on its natural enemies, particularly its parasitoids. For example, the overall performance of some parasitoids is severely reduced at high temperatures \citep{regniere2020modeling, hebert1990temperature, seehausen2017developmental, smith1986factors, royama2005analysis}, e.g.,  the SBW parasitoid \textit{Tranosema rostrale}, whose survival is negatively correlated with temperatures above 20 $^\circ$C \citep{regniere2020modeling,seehausen2017developmental}. Although we observe a high uncertainty for the threshold value of spring temperature, our results clearly illustrate the interaction dynamics between SBW and its parasitoids in relation to temperature and the consequences on defoliation levels.

The ACAR models perform quite well at both sites and reveal differences between the defoliation processes within each. There are almost 120 ecological sites in the reconstruction of \cite{navarro2018changes}; unfortunately, we  consider only two of these because of the unavailability of local climate data in the {\emph{NOAA}} (\url{https://www.ncei.noaa.gov/})  database for the full time period at the remaining sites. Fitting a classification and regression tree (CART) model \citep{loh2011classification} to all these sites could lead to a clustering analysis using the test statistic of Proposition \ref{pp:comparison} as a dissimilarity measure and therefore provide a deep dive into the SBW defoliation processes in Quebec's forests. 
A limitation of our study is that we rely on reconstructions of defoliation intensity. Thus, although we determine interesting insights into SBW-related defoliation, we clearly need direct observations of defoliation to validate our conclusions. For example, some prior distributions on model parameters could be extracted from our study to conduct Bayesian analysis on the shorter-length defoliation data (collected since the late 1960s) of the Ministère des Forêts, de la Faune et des Parcs (Quebec, Canada).

In regard to the statistical aspects, many questions remain about our ACAR model. For example, one can need a test for the condition of Proposition \ref{pp::solution}  to obtain a unique stationary and ergodic solution. 
Another statistically interesting question is the optimal selection of the set of covariates with penalized autoregression, like LASSO or Ridge penalties on the parameters $\gamma$ \citep{tibshirani1996regression, poignard2020asymptotic}.


\section*{Author contributions} 
Design and conceptualization: OJFO, ZMD, PM and MMG. R software programming: OJFO and ZMD. Models and data analysis: OJFO, ZMD and PM. Writing first draft: OJFO. Supervision and validation: PM and MMG. All authors provided input for the manuscript, contributed critically to the drafts and gave final approval for publication.

\section*{Acknowledgments} 

We acknowledge all the previous projects and researchers from which we compiled the dendrochronological data. 
We also thank Professor Paul Doukhan from Paris University for his valuable comments.

\section*{Funding} 

PM and MMG obtained funding from the Natural Sciences and Engineering Research Council of Canada (NSERC) Alliance and the Québec Ministry of Natural Resources and Forests (MRNF) to understand the dynamics of spruce budworm (ALLRP 558267-20). MMG obtained additional funding from an NSERC Discovery grant to reconstruct the regime of natural disturbances in forest ecosystems (RGPIN-2022-05423). ZMD received support from the Fonds de recherche du Qu\'{e}bec -- Nature et technologies (FRQNT) (PBEEE Scholarship DEBZI-334569) and a CY Initiative of Excellence (grant ”Investissements d’Avenir” ANR-16-IDEX-0008), Project ”EcoDep” PSI-AAP2020-0000000013.

\section*{Conflict of interest statement} 

The authors have no conflict of interest to declare.

\section*{Data availability statement} 

The defoliation data set is available on figshare  {\footnotesize \url{https://figshare.com/articles/dataset/Untitled_Item/7887746}.}

The underlying R code can be downloaded at \url{https://github.com/judiprodige/Adjacent-category-time-series-models.git}. A R package with ready-to-use functions will be available soon.

\newpage

\bibliography{biblio}

\newpage
\section*{Appendix}

\paragraph{Proof of Proposition \ref{pp::solution}}
Let us set $\lambda_t = (\lambda_{1,t}, \ldots, \lambda_{K,t})^\top.$ From Equation \eqref{eq::baseline_similarity}, 
$$
\lambda_t = \tilde \omega + \tilde A \overline Y_{t-1} + \tilde B \lambda_{t-1}, t \in \Z ,
$$
where $\tilde \omega = (\sum_{j = 1}^k \omega_j, {k=1, \ldots, K}), \tilde A = (\alpha|2\alpha|\cdots|K\alpha)^\top$ and 
$$
\tilde B = \begin{pmatrix}
    \beta_1 & 0 & \cdots & 0\\
    \beta_1 & \beta_2 & 0 \cdots \\
    \vdots &  & \ddots & \ddots  \\
    \beta_1 & \beta_2 & \cdots & \beta_K
\end{pmatrix}.
$$

The result of Proposition \ref{pp::solution} follows \citep{truquet2019coupling} Proposition 4 which provides stability conditions for category baseline multinomial autoregression. Indeed,    for $k=1,\ldots, K, |\beta_k|<1$, the roots $1/\beta_k, k = 1,\ldots, K$ of polynomial 
\begin{equation}
    \label{eq::cond_stationarity}
    \mathcal P(z) = \mathrm{det}\left(I - \tilde B z\right) = \Pi_{j=1}^K (1-\beta_j z)
\end{equation}
are outside the unit disk. Moreover, the moments of latent process depend on that of $(X_t)_{t\in Z}$, as $Y_{k,t}$'s take only two values : $0$ and $1.$

\paragraph{Condition for identifiability of } $\theta_0.$
\begin{enumerate}
    \item[{\bf I0}] For any $\theta \in  \Theta,~  |\beta_k|, k = 1, \ldots, K  < 1$ ; 
    \item[{\bf I1}] For any $v\in \R^P$,  
    $$v'X_1\in \mathcal{F}_0\vee \sigma(U_1)\Rightarrow v=0 ;$$ 
    \item[{\bf I2}] The distribution of $U_0$ is non-degenerate;
    \item[{\bf I3}] If $v$ is equal to any  $\alpha_{0,k}\iota, k = 1,\ldots,K$ or $\gamma_{0,p}\iota, p = 1,\ldots,P$ with $\iota = (1, \ldots, 1)^\top$ has $K$ components, the equalities $B^j v=B_0^j v,\quad j\geq 1$, entail $B=B_0$. 
\end{enumerate}
\subsection*{Derivative of latent process}

For $k= 1, \ldots, K,$

\begin{eqnarray}
\dpartial{\eta_{k,t}(\theta)}{\omega_\ell} & = & 0, \ell \neq k , \\
\dpartial{\eta_{k,t}(\theta)}{\omega_k} & =  & 1 + \beta_k  \dpartial{\eta_{k,t-1}(\theta)}{\omega_j} = 1 + \sum_{\ell \geq 1} \beta_k^\ell = \frac{1}{1-\beta_k} , \\
\dpartial{\eta_{k,t}(\theta)}{\gamma} & = & X_{t-1} + \beta_k \dpartial{\eta_{k,t-1}(\theta)}{\gamma} = \sum_{\ell \geq 0} \beta_k^\ell X_{t-\ell-1} ,
\\
\dpartial{\eta_{k,t}(\theta)}{\alpha} & = & \overline Y_{t-1} + \beta_k \dpartial{\eta_{k,t-1}(\theta)}{\alpha} = \sum_{\ell \geq 0} \beta_k^\ell \overline Y_{t-\ell-1} ,
\\
\dpartial{\eta_{k,t}(\theta)}{\beta_\ell} & = &  0, \ell \neq k , and \\
\dpartial{\eta_{k,t}(\theta)}{\beta_k} & = & \eta_{k, t-1}(\theta) + \beta_k \dpartial{\eta_{k,t-1}(\theta)}{\beta_k} = \sum_{\ell \geq 0} \beta_k^\ell \eta_{k, t-\ell-1}(\theta) .
\end{eqnarray}

\subsection*{Proof of Proposition \ref{pp::portemanteau}}
Let us denote by $\tilde \rho_h(\theta)$ the autocorrelation functions when the latent process is initialized at $t=0$ and  $\tilde \rho_h$ its value evaluate at $\theta = \theta_0$ and define $\rho_{1:q}(\theta)$ and $\tilde \rho_{1:q}(\theta)$ (resp. $\rho_{1:q}$ and $\tilde \rho_{1:q}$ accordingly. Under the conditions 
$$
n^{1/2}|\rho_{1:q} - \tilde \rho_{1:q} |_1 = o_\P(1); \quad \sup_\theta \| \nabla_\theta \rho_{1:q}(\theta) - \nabla_\theta \tilde \rho_{1:q}(\theta) \| = o_\P(1) \text{ and } \lim_{n\to\infty} \E \sup_\theta \|\nabla_\theta^2  \rho_{1:q}(\theta) \| < \infty 
$$
obtained with $\E|X_0|_2^r < \infty$ for sustainable $r,$

\begin{equation}
    \label{eq::hatrho}
    \sqrt{n}\hat\rho_{1:q} = \sqrt{n}\rho_{1:q} + C_q \sqrt{n}(\hat\theta - \theta_0) + o_\P(1)
\end{equation}

see \cite{francq2019garch}, proof of Theorem 8.2 for the case of GARCH model.
From Proposition \ref{pp::asymptotics},
$$
\sqrt{n}(\hat \theta_n - \theta_0 ) = -J^{-1}\frac{1}{\sqrt{n}}\sum_{t=1}^n s_t(\theta_0)  + o_\P(1).
$$
We also have 
\begin{equation}
    \label{eq::iota}
    \sqrt{n}\rho_{1:q} =   n^{-1/2}\sum_{t=1}^n \sum_{k = 1}^K e_{k,t} \iota_k.
\end{equation}

It follows that 
$$
\sqrt{n}\begin{pmatrix}
    \hat \theta_n - \theta_0 \\
    \rho_{1:q}
\end{pmatrix}
= \frac{1}{\sqrt{n}}\sum_{t=1}^n \sum_{k = 1}^K e_{k,t}
\begin{pmatrix}
    J^{-1} \nabla_{\theta_0}\eta_{k,t}(\theta) \\
    \iota_{k,t} 
\end{pmatrix} + 0_\P(1).
$$
Because the sequence $\left(\sum_{k = 1}^K e_{k,t}
\begin{pmatrix}
     \nabla_{\theta_0}\eta_{k,t}(\theta) \\
    \iota_{k,t}
\end{pmatrix}\right)_{t\in \Z}$
is a martingale difference sequence,
$$
\sqrt{n}\begin{pmatrix}
    \hat \theta_n - \theta_0 \\
    \rho_{1:q}
\end{pmatrix} \Rightarrow \mathcal N\left(0_{Kq+Q}, \begin{pmatrix}
    J^{-1} L J^{-1^\top} & G^\top\\
     G & D
\end{pmatrix}\right).
$$
From Equation \eqref{eq::hatrho}, 
$$
\sqrt{n}\hat  \rho_{1:q}  \Rightarrow \mathcal N(0_{Kq}, D +   C_q   J^{-1}   L  {  J^{-1}}^\top   C_q^\top +   G   C_q^\top +   C_q    G^\top).
$$

Under \ref{ass::invertible} , $$ D +   C_q   J^{-1}   L {  J^{-1}}^\top   C_q^\top +   G   C_q^\top +   C_q    G^\top  = \E ZZ^\top$$ with  
$$
Z = \sum_{k=1}^K e_{k,0}\left(\iota_k + C_q J^{-1} \nabla_{\theta_0}\eta_{k,0}(\theta) \right)
$$ is invertible. Then, 
$$
n \hat \rho_{1:q}^\top\left(D +   C_q   J^{-1}   L  {  J^{-1}}^\top   C_q^\top +   G   C_q^\top +   C_q    G^\top \right)^{-1} \hat  \rho_{1:q}  \Rightarrow  \chi_{Kq}^2.
$$
The result of the proposition follows from continuous map theorem.

\subsection*{Proof of Proposition \ref{pp:comparison}}

From Proposition \ref{pp::asymptotics}, $ \begin{pmatrix}
    s_t^{(1)}(\theta_0^{(1)}) \\
    s_t^{(2)}(\theta_0^{(2)})
\end{pmatrix}_{t\in\Z}$  is a  square integrable martingale difference sequence. Then, 

\begin{eqnarray}
 \sqrt{n}
\begin{pmatrix}
    \hat \theta_n^{(1)} -  \theta_0^{(1)} \\
    \hat \theta_n^{(2)} -  \theta_0^{(2)}
\end{pmatrix}
 & = & -\frac{1}{\sqrt{n}} \sum_{t=1}^n \begin{pmatrix}
    J_1 ^ {-1} s_t^{(1)}(\theta_0^{(1)}) \\
    J_2 ^ {-1} s_t^{(2)}(\theta_0^{(2)})
\end{pmatrix} + 0_\P(1) \\
& \Rightarrow  & \mathcal N\left(0_{2(3K+P)}, 
\begin{pmatrix}
   J_1^{-1}L_1{J_1^{-1}}^\top & {J_1^{-1}} S^\top  {J_2^{-1}}^\top\\
    J_2^{-1}  S {J_1^{-1}}^\top & J_2^{-1}L_2{J_2^{-1}}^\top
\end{pmatrix} 
\right).
\end{eqnarray}

 It follows that, under the null hypothesis, $$\sqrt{n}\left(\theta_n^{(1)} - \theta_n^{(2)} \right) = \sqrt{n}\left(\hat \theta_n^{(1)} -  \theta_0^{(1)}\right) - \sqrt{n}\left(\hat \theta_n^{(2)} -  \theta_0^{(2)}\right) \Rightarrow \mathcal N (0, V)$$ with 
 $$
 V = J_1^{-1}L_1{J_1^{-1}}^\top + J_2^{-1}L_2{J_2^{-1}}^\top + J_2^{-1}  S {J_1^{-1}}^\top  + {J_1^{-1}} S^\top  {J_2^{-1}}^\top = \E UU^\top ,
 $$
 and $U =  J_1^{-1}s_t^{(1)}(\theta_0^{(1)}) +  J_1^{-2}s_t^{(2)}(\theta_0^{(2)}).$
The matrix $V$ is then invertible under \ref{ass::invertible2}.
The result of the proposition follows from the continuous map theorem.

\paragraph{Threshold estimate for temperature effect}

The thresholds are derived using the delta method \citep{van2000asymptotic}. Denoting $\theta_0(x)$ the regression parameter of covariate $x,$ the threshold of the temperature effect in spring, for example, is $\mathrm{TS} = \frac{\theta_0(\Delta \text{temp. max. spring})}{2 \times \theta_0(\text{Square of} \Delta\text{temp. max. spring})}.$ Then, $\widehat{\mathrm{TS}}$ is obtained by replacing $\theta_0(x)$ by $\hat \theta(x).$ The variance of $\widehat{\mathrm{TS}} - \mathrm{TS}$ is $\hat \mu^\top \hat \Sigma_{spring}\hat \mu$ , where $$\hat \mu^\top = \left(\frac{1}{2 \times \hat \theta(\text{Square of} \Delta\text{temp. max. spring})}, - \frac{\hat \theta(\Delta \text{temp. max. spring})}{2 \times \hat \theta^2(\text{Square of} \Delta\text{temp. max. spring})}\right) , $$
and $\hat \Sigma_{spring} $ is the estimate covariance matrix of $$(\hat \theta(\Delta \text{temp. max. spring}), \hat \theta(\text{Square of} \Delta\text{temp. max. spring})).$$

\begin{table}[!ht]
    \tiny
    \centering
    \caption{
    Results of the quasi-maximum likelihood estimation}\label{tab::simulations}
    \begin{tabular}{|l|r|cccccccccccccc|}
    \cline{2-16}
  \multicolumn{1}{l|}{\emph{n}}  & $\theta_0$ & 1.2 & 0.7 & 0.5 & -0.8 & 1.5 &  -1.5 & 2.0 & 2.0 & 0.3 & -0.3 & 0.5 & 0.8 & -0.2  & 0.3 \\ \hline
  70 & CMLE &  3.616 &   3.311 &   1.687 &  -1.613 &   3.144 &  -3.108 &   4.252 &   4.165 &  -0.583 &  -1.507 & -0.304 &   0.818 &  -0.132 &   0.298  \\
  & TSE & 6.794 &  7.424 &  7.080 &  0.735 &  1.246 &  1.223 & 1.577 &  1.594 &  7.216 & 7.706 & 
 7.425 &  0.099 &  0.435 &  0.345 \\
  &  MAE & 3.482 & 3.985 &  3.058 &  2.154 &  2.716 &  3.597 &  3.773 & 3.692 &  3.208 &  3.912 &  3.659 & 0.824 &  0.970 &  0.806 \\
  & MSE & 8.783 &  10.275 &  9.006 &   2.924 &  4.256 &   4.895 &   6.040 &   5.579 &   9.017 &   9.459 & 10.067 &  1.032  &  1.175  &  1.008 \\  \hline
 \rowcolor[gray]{.9}100 & CMLE & 1.642  & 0.854 &  0.625  & -1.097 &  1.908 &  -1.966 &  2.581  & 2.577 & 0.180 & -0.709 &  0.649  & 0.809 &  -0.171 &  0.273 \\
\rowcolor[gray]{.9}  & TSE  & 1.011 & 1.317 & 1.130 & 0.400 & 0.588 & 0.622 & 0.784 & 0.774 & 1.203 & 1.346 & 1.314 & 0.064 & 0.337 & 0.292 \\
 \rowcolor[gray]{.9}  & MAE & 0.756 & 0.641 & 0.646 & 1.200 & 0.925 & 2.122 & 1.258 &  1.258 & 0.903 &  1.646 &  0.819 & 0.736 & 0.537 &  0.097 \\
\rowcolor[gray]{.9}   & MSE & 1.666 & 1.809  & 1.835 &  0.219 & 0.313  & 0.345 &  0.470 &  0.451 & 1.828 & 1.819 & 1.863 & 0.034 & 0.129 &  0.130 \\ \hline
   300 & CMLE & 1.237 & 0.636 & 0.580 & -0.805 & 1.492 & -1.531 & 1.997 & 2.012 &  0.250 & -0.394 & 0.460 & 0.807 & -0.207 & 0.308 \\
   & TSE & 0.359 & 0.520 &  0.404 & 0.179 &  0.277 & 0.284 & 0.356 & 0.358 & 0.432 & 0.510 & 0.494  & 0.035  & 0.192 & 0.158\\
    & MAE &  0.843  & 0.191 & 0.123 &  0.979 &  1.223 &  1.784  & 1.644 &  1.642 &  0.134 &  0.810 & 0.103 &  0.765 & 0.392 & 0.143 \\
    & MSE & 0.122 & 0.100 & 0.114 & 0.051 &  0.089 &  0.097 & 0.123 &  0.117 &  0.130  & 0.122 & 0.153 & 0.011 &  0.049  & 0.042 \\ \hline
 \rowcolor[gray]{.9}500 & CMLE & 1.182 & 0.578 & 0.518 & -0.801 &  1.486 & -1.492 &  1.973 & 1.966 & 0.277 & -0.292 & 0.539 & 0.799 &  -0.207  & 0.291 \\
\rowcolor[gray]{.9} & TSE & 0.265 & 0.392 & 0.307 & 0.137 & 0.216 & 0.216 & 0.276 &  0.275 & 0.329 & 0.376 & 0.373 & 0.028  & 0.152 &  0.127 \\
\rowcolor[gray]{.9}  & MAE &  0.935 &  0.308 &  0.193 & 0.937  & 1.284 & 1.716 &  1.724 &  1.725  & 0.060 & 0.676 &  0.137 & 0.772  & 0.352 & 0.173 \\
\rowcolor[gray]{.9} & MSE &  0.064 &  0.061 &  0.067 &  0.032 &  0.057 &  0.053  & 0.072 & 0.071 &  0.075 & 0.077 & 0.089  & 0.007 & 0.032 &  0.028 \\
 \cline{2-16}
     & $\theta_0$ & 1.2  & 0.7  & 0.5  & 0.8 & -1.5 &  1.5 & -2.0  & -2.0 &  -0.3 &   0.3 &  -0.5 & -0.8 &  0.2 &  -0.3 \\ \hline
     70 & CMLE & 4.484 &   1.885 &  -0.373 &   2.239 &  -4.037 &   4.062 &  -5.408 &  -5.373 &  -0.289 &   1.602 &   -0.402 &  -0.804 &   0.116 &  -0.316 \\
     & TSE & 2.937 &  1.474 &  1.459 &  0.761 &  1.290 &  1.265 &  1.638 &  1.668 &  1.543 & 2.348 & 
 1.500 &  0.061 &  0.376 &  0.329 \\
 & MAE & 5.424 &  2.649 &  2.531 &  2.486 &  3.922 & 4.215 &  5.270 &  5.232 & 2.800 &  3.881 &  2.892 & 1.111 &  0.919 &  0.938 \\
 & MSE & 11.246 &   4.721 &   5.134 &  4.489 &   6.677 &   6.917 &   8.772 &   8.434 &  5.974 &  9.647 &  6.247 & 
 1.271 &  1.137 &   1.121 \\  \hline
  \rowcolor[gray]{.9} 100 & CMLE & 2.286 & 1.077 & 0.401 & 1.237 & -2.345 & 2.356 & -3.117 & -3.186 & -0.305 & 1.209 & -0.686 & -0.803 & 0.157 & -0.298 \\
 \rowcolor[gray]{.9}    & TSE & 1.956 & 0.918 &  0.986 & 0.451 & 0.718 & 0.738 & 0.925 & 0.941 & 0.897 & 1.403 & 0.926 & 0.050 & 0.319 & 0.303 \\
 \rowcolor[gray]{.9}     & MAE & 1.808 & 0.889 & 0.956 & 0.547 & 0.932 & 0.946 & 1.205 & 1.274 & 0.842 & 1.492 & 1.111 & 0.043 & 0.151 &  0.127 \\
 \rowcolor[gray]{.9}    & MSE & 3.910 &  1.730 &  1.603 &  1.373 & 2.356 & 2.377 & 3.276 &  3.578 & 1.256 & 5.799 &  2.350 
 & 0.058 & 0.198  & 0.166 \\ \hline
  300 & CMLE &  1.171 & 0.638  & 0.401 & 0.849 & -1.611 & 1.600 & -2.123 & -2.112 & -0.198 & 0.490 & -0.352 & -0.808 &  0.189 & -0.319 \\
 & TSE &  0.803 & 0.378 & 0.480 & 0.171 & 0.273 & 0.271 & 0.348 & 0.347 & 0.365 & 0.438 & 0.360 &  0.028 & 0.184 & 0.167 \\
 & MAE & 0.651 & 0.349 & 0.393 & 0.124 & 0.192 & 0.175 & 0.234 & 0.245 & 0.378 & 0.455 & 0.373 & 0.020 & 0.072 & 0.073 \\
 & MSE & 0.845 & 0.461 & 0.513 & 0.162 & 0.260 & 0.252 & 0.318 & 0.324 & 0.472 & 0.578 & 0.459  & 0.026 & 0.092 & 0.094 \\ \hline
 \rowcolor[gray]{.9}  500 & CMLE & 1.143 &  0.654 & 0.461 & 0.821 & -1.538 & 1.556  & -2.053  & -2.065 & -0.196 & 0.455 & -0.389 & -0.806 & 0.184 & -0.310\\
\rowcolor[gray]{.9}  & TSE & 0.607 & 0.282 & 0.360  & 0.127 &  0.203 & 0.205 & 0.262 & 0.263 & 0.274  & 0.320 & 0.265 & 0.021 &  0.144 & 0.131 \\
 \rowcolor[gray]{.9}  & MAE &0.601 &  0.266 &  0.280 &  0.093 &  0.136 &  0.155 & 0.176 & 0.190 & 0.268 & 0.346 & 0.264 & 0.018 & 0.056 & 0.053 \\
 \rowcolor[gray]{.9}  & MSE &  0.738 &  0.328 & 0.365 & 0.119 & 0.180 & 0.194 & 0.224 & 0.247 & 0.340 & 0.462 & 0.326 & 0.021 & 0.072 & 0.065\\ \cline{2-16}
 & $\theta_0$  & 1.2 &  0.7  & 1.5 &  0.8 & -1.5 & -1.5 &  2.0 & -2.0  & 0.3 & -0.3 & -0.5 & -0.8 &  0.2 &  -0.3 \\
\hline 
70 & CMLE &  6.218 &  2.849 &   2.043 &  3.363 &  -6.405 &  -6.394 &   8.474 &  -8.528 &   2.311 & 0.434 &  0.235 &  -0.804 &  0.0742 &  -0.310 \\
 & TSE & 3.610 &  1.817 &  1.786 &  0.968 &  1.551 &  1.603 &  2.042 &  2.103 & 2.1483 &  4.074  &  1.733 &  0.060 &  0.383 & 0.352 \\
 & MAE & 7.506 &  4.794 &  3.945 &  3.563 &  6.421 &  6.409 &  8.480 &  8.530 &  4.885 & 6.469 &  4.185 &  1.179 & 
1.002 &  1.033\\
 & MSE & 14.640 &   9.501 &   7.536 &   6.173 &  10.750 &  10.669 &  13.978 & 14.160 &  9.441 &  16.167 &  7.893 & 
  1.415 &   1.196 &  1.234 \\ \hline
\rowcolor[gray]{.9} 100 & CMLE & 1.870 & 0.861 &  1.385 &  1.249 &  -2.278  & -2.265 &  2.965 & -3.053 & 0.852  & 0.041 & -0.284 & -0.805 & 0.138 & -0.319 \\ 
\rowcolor[gray]{.9} & TSE & 1.834 & 0.993 & 1.142 &  0.457 & 0.720 & 0.728 & 0.919 & 0.950 & 1.017 & 1.517 & 0.866 & 0.050 & 0.359 & 0.304 \\
\rowcolor[gray]{.9} & MAE & 1.542 & 0.886 & 0.922 & 0.524 & 0.837 & 0.828 & 1.025 & 1.107 & 1.101 & 1.413 & 0.950 & 0.043 & 0.162 & 0.134 \\
\rowcolor[gray]{.9} & MSE &  2.017 & 1.142 & 1.275 & 0.848 & 1.421  & 1.467 & 1.697 & 1.971 & 1.608 & 2.059 & 1.329 &  0.057 & 0.221 & 0.171 \\  \hline
300 & CMLE & 1.324 & 0.795 & 1.448 &  0.874 & -1.653 & -1.650 & 2.196 & -2.218  & 0.378 & -0.247 & -0.468 & -0.802 & 0.183 & -0.308 \\
& TSE & 0.853 &  0.438 & 0.545 & 0.181 & 0.293 &  0.292 & 0.379 & 0.381 & 0.378 & 0.499 & 0.371 & 0.028 & 0.192 & 0.181 \\
& MAE &  0.717 & 0.336 & 0.393 & 0.138 & 0.239 & 0.229 & 0.294 & 0.313 & 0.388 & 0.508 & 0.336 & 0.021 & 0.077 &  0.065 \\
& MSE &  0.861 & 0.451 & 0.504 & 0.188 & 0.332 & 0.311 & 0.406 & 0.418 & 0.490 & 0.647 & 0.422 & 0.027 & 0.096 & 0.083 \\  \hline
\rowcolor[gray]{.9}500 & CMLE & 1.199 & 0.718 & 1.458 &  0.842 &  -1.580 & -1.587 & 2.109 & -2.116 &  0.380 
 & -0.207 & -0.431 & -0.803 & 0.183 & -0.307 \\
\rowcolor[gray]{.9} & TSE & 0.647 & 0.325 & 0.418 & 0.132 & 0.214 & 0.216 & 0.279 & 0.280 & 0.286 & 0.373 & 0.274 & 0.022 & 0.151  & 0.140 \\ 
\rowcolor[gray]{.9} & MAE & 0.537 & 0.298 & 0.353 & 0.103 & 0.161 & 0.168 & 0.207 & 0.201 & 0.313 & 0.364 & 0.264 &  0.017 &  0.060 & 0.051 \\
\rowcolor[gray]{.9} & MSE  & 0.671 & 0.391 & 0.442 & 0.135 & 0.213 & 0.218 & 0.273 & 0.267 & 0.405 & 0.468 & 0.345 & 0.021 & 0.074 & 0.064 \\ \hline
    \end{tabular}
    \end{table}

\end{document}